%Paper: gr-qc/9404019
%From: Eric Woolgar <woolgar@cs.dal.ca>
%Date: Mon, 11 Apr 1994 19:45:16 -0300

	     \magnification=1200
	%     \voffset  = 20 true mm  %space between top of page and text
	%     \hoffset  = 28 true mm  %left margin
	     \hsize    = 165 true mm %length of line
	     \vsize    = 220 true mm  %vertical length of text
	     \parskip  = 3 true  pt plus 1 true pt minus 1 true  pt
	%     3 pt extra between paragraphs
		\baselineskip = 16 true pt plus 1 true pt minus 1 true pt

	\def\centreline{\centerline}

	%%%% number 0 %%%%

	\def\dal{\displaystyle{{\hbox to 0pt{$\sqcup$\hss}}\sqcap}}

	\noindent
	{\centreline {\bf The Positivity of Energy for
	Asymptotically Anti-de Sitter Spacetimes}
	\vskip 1 true cm
	\centreline{E. Woolgar}
	\centreline{Dept. of Mathematics, Statistics, and Computer Science}
	\centreline{Dalhousie University}
	\centreline{Halifax, Nova Scotia}
	\centreline{Canada  B3H 3J5}
	\vskip 2 true cm
	{\centreline {\bf Abstract}}
	\vskip 0.5 true cm
	\noindent
	We use the formulation of asymptotically anti-de Sitter boundary conditions
	given by Ashtekar and Magnon to obtain a coordinate expression for the
	general asymptotically AdeS metric in a neighbourhood of infinity. From this,
	we are able to compute the time delay of null curves
	propagating near infinity. If the gravitational
	mass is negative, so will be the time delay (relative to null geodesics
	at infinity) for certain null geodesics in the spacetime.
	Following closely an argument given
	by Penrose, Sorkin, and Woolgar, who treated the asymptotically flat case,
	we are then able
	to argue that a negative time delay is inconsistent with
	non-negative matter-energies in spacetimes having good causal properties.
	We thereby obtain
	a new positive mass theorem for these spacetimes. The theorem may be
	applied even when the matter flux near the boundary-at-infinity falls off
	so slowly that the mass changes, provided the theorem is applied in a
	time-averaged sense. The theorem also applies in certain spacetimes having
	local matter-energy that is sometimes negative, as can be the case in
	semi-classical gravity.
	\par\vfil\eject
	\noindent
	\centreline{\bf The Positivity of Energy for Asymptotically Anti-de Sitter
	Spacetimes}
	\vskip 1 true cm
	\noindent
	{\centreline {\bf Introduction}}
	\vskip 0.5 true cm
	\noindent
	Although most studies of the asymptotic properties of spacetime have
considered
	the asymptotically flat case, there are several reasons to study spacetimes
	that asymptote to constant non-zero curvature manifolds, and in particular
	to the negative curvature case. The first reason is that the analysis is
	generally simpler. As we shall see, the boundary-at-infinity
	comes in one piece, unlike the asymptotically flat case with its two null
	pieces joined at the singular point $i^0$ representing spatial infinity.
	Furthermore, the negative cosmological constant acts as a mass term in the
	graviton equations of motion, so it is natural to impose boundary conditions
	that demand there be no gravitational radiation flux at infinity; in turn,
this
	greatly simplifies the asymptotic metric.

	An additional reason for studying
	asymptotic behaviour of this type is provided by
	supergravity. Certain physically interesting supergravity models require
	a negative cosmological constant, while other related models have
	scalar fields whose contribution to the energy can be negative.
	These models do not admit flat space as a solution (except when the scalar
	field potential vanishes);
	rather, anti-de Sitter space is the natural classical ground state
	(lowest energy classical solution) of these models. It is reasonable that
	anti-de Sitter space be a ground state --- in particular, it has vanishing
	Weyl curvature, and mass-energy would be expected to
	vanish in conformally flat spacetimes.
	However, in the case of the supergravities with the scalar fields present,
	it was not known if the hamiltonians were bounded below.
	This served as motivation for the study of the stability of these models
	during the early 1980s. It was eventually shown that the dynamics of these
	theories implied that when the scalar field potentials were negative, the
	other fields were driven to positive energy configurations, such that the net
	energy was bounded below,
	provided the fields obeyed suitable boundary conditions.
	This result first emerged in a small perturbations analysis of
	Breitenlohner and Freedman.$^{(1)}$
	Abbott and Deser$^{(2)}$ then gave a positive energy
	argument, based on supergravity but applied to pure general relativity,
	implying that the energy of anti-de Sitter space
	was a true lower bound in general relativity for all spaces obeying the
	appropriate boundary conditions. A full Witten-type$^{(3)}$ proof for
	general relativity
	was provided by Gibbons, Hull, and Warner$^{(4)}$, who also addressed the
	question of positivity of energy in the supergravity models.
	The issue of exactly what
	constituted ``suitable boundary
	conditions'' was
	resolved by Hawking.$^{(5)}$ Ashtekar and Magnon$^{(6)}$
	reformulated these conditions and
	investigated the asymptotic properties of spacetimes obeying them.

	Herein we adopt the Ashtekar and Magnon formulation, and derive from it
	a completely general coordinate form for the spacetime metric on a global
	neighbourhood of infinity. We then use this form of the metric to obtain
	a positive mass theorem that has no direct relation to supergravity (at
	least, none that is apparent), and
	in particular does not use spinorial methods in its derivation. Rather,
	we adapt to the present case
	the method of Penrose, Sorkin, and Woolgar,$^{(7)}$
	who recently obtained
	a positivity theorem for the mass of asymptotically flat spacetimes.
	This method depends on an analysis of the time delay effect for null
	geodesics moving under the influence of a gravitating mass. We show that
	the time delay due to a negative mass is incompatible with the geometric
	focussing effect ({\it i.e.} the gravitational lens effect, if one prefers) of
	generic curvature created by positive local matter-energy distributions.

	In part, our motivation is to present the method of ref. (7), suitably
	modified, in a context which has fewer complications than are present in
	the asymptotically flat case. Secondly, it now appears that the method
	of ref. (7), suitably generalised,$^{(8)}$
	may actually yield somewhat different information in
	certain contexts than can be extracted from the spinorial proofs,
	leading to new Bogomol'nyi-type bounds on the mass and charges of
	fields. There is a third motivation. This treatment of anti-de Sitter
	space is a natural step in a programme that should next be directed to the
	study of asymptotically Robertson-Walker metrics, especially those with
	positive spatial curvature. It is remarkable that
	very little analytical work exists in the general Robertson-Walker
	case, considering especially
	that perturbations of FRW and de Sitter spacetimes are so important in
	the study of physical cosmology. For further motivation along similar
	lines, see ref. (9).

	We follow the conventions of ref. (10), and use early latin letters as
	abstract indices, while greek indices refer to coordinate systems, and
	middle latin indices refer to certain directions within those coordinate
	systems. For a weaker version of the argument herein, which applies only
	to asymptotically AdeS-Schwarzschild spacetimes, see ref. (11).

	\vskip 0.5 true cm
	\noindent
	{\centreline{\bf Asymptotic Coordinate System}}
	\vskip 0.5 true cm
	\noindent
	Here we restate what is sometimes known as
	Hawking's reflexive boundary condition,$^{(5)}$ in the formulation of
	Ashtekar and Magnon.$^{(6)}$

	{\smallskip\narrower\noindent
	{\underbar{Definition 1:}} A spacetime $({\cal M},g_{ab})$ is
	asymptotically anti-de Sitter if there exists a manifold-with-boundary
	${\tilde {\cal M}}$, a $C^3$
	metric ${\tilde g}_{ab}$ on ${\tilde {\cal M}}$,
	and a diffeomorphism from ${\cal M}$ onto ${\tilde {\cal M}}\backslash
	\partial {\tilde {\cal M}}$ such that
	\item{(i)}{there exists a $C^3$
	function $\Omega$ on ${\tilde {\cal M}}$
	such that ${\tilde g}_{ab}=\Omega^2 g_{ab}$ on ${\cal M}$,}
	\item{(ii)}{${\cal I}=\partial {\tilde {\cal M}}$ has topology $S^2\times R$
	and is defined as the surface $\Omega=0$,}
	\item{(iii)}{$g_{ab}$ satisfies $R_{ab}-{1\over 2}g_{ab}R+\Lambda g_{ab}
	=8\pi T_{ab}$, with $\Lambda<0$ (we choose units in which $\Lambda=-3$),
	where $\Omega^{-3}T_a^b$ admits a smooth limit to ${\cal I}$, and}
	\item{(iv)}{the group of conformal isometries of $({\cal I},q_{ab})$ is the
	anti-de Sitter group (or covering group thereof), $q_{ab}$ being the
	restriction of ${\tilde g}_{ab}$ to ${\cal I}$.}
	\smallskip}

	\noindent
	Condition (ii) shows that we are considering spaces that are asymptotic to
	{\it universal} anti-de Sitter space, which is $R^4$ --- often the time axis
	is quotiented by the integers to produce a spacetime with closed
	timelike curves. Although we will work with Definition 1, the positivity
	theorem we obtain will hold under the quotient. Fig. (1) shows (universal)
	AdeS spacetime embedded in the Einstein cylinder, which is a prototype
	for the above embedding. For another prototypical case, see the description
	of the AdeS-Schwarzschild solution given in Appendix 2.

	In contrast with the asymptotically flat case,$^{(12)}$ the smoothness
	conditions in this definition are probably appropriate. Because the
	cosmological constant affects small perturbations of the metric in the manner
	that a mass term would, it is reasonable to believe that outgoing radiation
	does not reach ${\cal I}$, so smooth generic initial data are unlikely
	to destroy smoothness at ${\cal I}$
	at later times (although the author is not aware
	of specific results in the AdeS case). It is therefore likely that $C^3$
	smoothness at ${\cal I}$
	is a consistent assumption. Furthermore, it seems a necessary
	assumption
	because ${\cal I}$ is timelike (as we shall verify), and so
	inbound radiation flux at ${\cal I}$ is possible.
	Thus, asymptotically AdeS spaces are not globally
	hyperbolic, and would be unsuitable for many physical calculations unless
	the boundary conditions forbid such flux. Ref. (6) shows
	that the absence of Bondi flux at ${\cal I}$ is derivable from the above
	smoothness conditions.

	Condition (iv) permits us to write the line element of $q_{ab}$ as
	$$-dt^2+d\theta^2+\sin^2\theta d\phi^2\quad .\eqno{(1)}$$
	We extend these coordinates to a global neighbourhood of ${\cal I}$ in any
	convenient way.
	The ``physical'' line element corresponding to $g_{ab}$ will be conformal
	to the warped product
	$$-dt^2+d\theta^2+\sin^2\theta d\phi^2
	+\alpha d\xi^2 +{\cal O}(f(\xi))\quad ,\eqno{(2)}$$
	where $f(\xi)$ vanishes on ${\cal I}$. We use ``${\hat =}$'' to denote
	equality on ${\cal I}$, so the statement that $f$ vanishes on ${\cal I}$
	is written as $f(\xi){\hat =}0$.
	Here there are no cross-terms ($d\xi dt$, $d\xi d\theta$, or $d\xi d\phi$)
	on ${\cal I}$ because we define $\xi$ such that ${\cal I}$
	is a level surface on which $t$, $\theta$, and $\phi$ serve as
	coordinates --- if ${{\partial}\over {\partial \xi}}$ is ``metric-dual'' to
	$d\xi$, it must then be orthogonal to ${\cal I}$.

	Without loss of generality, we can thus require that the physical line element
	be conformal to
	$$-dt^2 + (1-\Omega^2)(d\theta^2+\sin^2\theta d\phi^2)+\beta{{d\Omega^2}\over
	{(1-\Omega^2)}} +{\cal O}(\Omega)\quad ,\eqno{(3)}$$
	where $\Omega=\Omega(\xi){\hat =}0$ and where $\beta(\Omega)=\alpha(\xi)$.
	In particular, we take $\Omega=\cos\xi$,
	with ${\cal I}$ defined by $\xi={{\pi}\over 2}$.
	Invoking (i) and (ii) above gives the
	physical line element
	$$ds^2=\Omega^{-2}\bigg ( -dt^2 +\beta{{d\Omega^2}\over {(1-\Omega^2)}}
	+(1-\Omega^2)(d\theta^2+\sin^2\theta d\phi^2)+{\cal O}(\Omega)\bigg )
	\quad .\eqno{(4)}$$

	Next, let us write the Einstein equation in terms of the quantity
	$$S_{ab}=R_{ab}-{1\over 6}g_{ab}R\quad ,\eqno{(5)}$$
	since $S_{ab}$ has somewhat simpler properties under conformal transformations
	than the Einstein tensor. The Einstein equation with cosmological constant
	(normalised so that $\Lambda=-3$; this can always be achieved by rescaling
	the coordinates) can be written as
	$$S_{ab}+g_{ab}=8\pi g_{ac}\big ( T^c_b-{1\over 3}\delta^c_b T\big )
	=8\pi g_{ac}{\cal T}_b^c\quad .
	\eqno{(6)}$$
	Let us now define ${\tilde S}_{ab}$ to be given from equation (5), but with
	$g_{ab}$ and its Ricci tensor replaced by ${\tilde g}_{ab}$ and its Ricci
	tensor.
	The conformal transformation property alluded to above is
	$$S_{ab}={\tilde S}_{ab}+{2\over {\Omega}}{\tilde \nabla}_a{\tilde \nabla}_b
	\Omega -{1\over {\Omega^2}}{\tilde g}_{ab}{\tilde g}^{cd} {\tilde \nabla}_c
	\Omega {\tilde \nabla}_d\Omega\quad ,\eqno{(7)}$$
	where ${\tilde \nabla}_a{\tilde g}_{bc}=0$. Hence the field equation becomes
	$${\tilde S}_{ab}+{1\over {\Omega^2}}{\tilde g}_{ab}\big ( 1-
	{\tilde g}^{cd}{\tilde \nabla}_c\Omega{\tilde\nabla}_d\Omega\big ) +{2\over
	{\Omega}}{\tilde\nabla}_a{\tilde\nabla}_b\Omega={{8\pi}\over {\Omega^2}}
	{\tilde g}_{ac}{\cal T}^c_b \quad .\eqno{(8)}$$

	Given the smoothness requirements on ${\tilde g}_{ab}$, then ${\tilde S}_{ab}$
	is finite at ${\cal I}$ and, given the fall-off condition (iii) on the stress
	tensor, the right-hand-side of equation (8) is ${\cal O}(\Omega)$, so we
	obtain from the field equation (8)
	the usual result$^{(13)}$ that ${\cal I}$ has spacelike normal $n_a={\tilde
	\nabla}_a\Omega$.
	$$ {\tilde g}^{ab}{\tilde\nabla}_a\Omega{\tilde\nabla}_b\Omega = {\tilde
	g}^{ab}n_an_b{\hat =}1\quad .\eqno{(9)}$$
	 From this, we get that ${\tilde
	g}_{11}{\hat =}\beta{\hat =}1$, since ${\tilde g}_{1i}{\hat =}0$, where from
	here onward $i\in \{ 0,2,3\}$ denotes the coordinates $(t,\theta,\phi)$.
	All this is of course standard,$^{(14)}$
	including the next step, which is to use the gauge
	freedom available in the choice of $\Omega$ to set
	$$\Omega^{-1}\big ( {\tilde g}^{ab}n_an_b-1\big ) {\hat=}0\quad ,\eqno{(10)}$$
	whence the field equation, the differentiability of ${\tilde g}_{ab}$, and
	the fall-off condition (iii) imply that
	$${\tilde\nabla}_a{\tilde\nabla}_b \Omega{\hat =}0\quad .\eqno{(11)}$$
	This can also be written in our coordinates as ${\tilde \Gamma}^1_{ab}{\hat =}
	0$, and it implies that
	${\tilde g}_{11,1}{\hat =}0$ and ${\tilde g}_{ij,1}{\hat =}0$,
	so we have eliminated
	all ${\cal O}(\Omega)$ terms except those in ${\tilde g}_{1i}$. However,
	the coordinate transformation
	$$\eqalign{x^i\longrightarrow{\bar x}^i=&x^i+\Omega^2f^i(x^j)\quad ,\cr
	dx^i\longrightarrow d{\bar x}^i=&dx^i+2\Omega f^i(x^j) d\Omega +{\cal O}
	(\Omega^2)\quad ,\cr
	\Rightarrow
	d{\bar x}^i d{\bar x}^j=& dx^idx^j+4\Omega f^{(i}dx^{j)} + {\cal O}(\Omega^2)
	\quad ,\cr}\eqno{(11)}$$
	will remove any such terms at the expense only of new terms of ${\cal O}
	(\Omega^2)$, and so we may choose the coordinates so no ${\cal O}(\Omega)$
	terms appear in the conformal metric, which takes the general form
	$$\eqalign{d{\tilde s}^2=& -dt^2+{{d\Omega^2}\over {(1-\Omega^2)}} +(1-
	\Omega^2)(d\theta^2+\sin^2\theta d\phi^2)\cr
	&+\Omega^2 \big ( Adt^2+Bd\Omega^2+Cd\theta^2+D\sin^2\theta d\phi^2\big )\cr
	&+2\Omega^2\big ( L dt d\theta + O dtd\phi+Gd\theta d\phi\big )\cr
	&+2\Omega^2\big ( Xdt+Yd\theta+Zd\phi\big ) d\Omega\cr
	&+\Omega^3\big ( Jdt^2+Kd\Omega^2+Nd\theta^2+F\sin^2\theta d\phi^2\big )\cr
	&+2\Omega^3\big ( Mdtd\theta +Sdtd\phi +Qd\theta d\phi\big )\cr
	&+2\Omega^3\big ( Udt+Vd\theta+Wd\phi\big )d\Omega+{\cal O}(\Omega^4)\quad
.\cr}
	\eqno{(12)}$$

	The coefficients $B$ and $K$ in the above metric can be cancelled by the
	coordinate transformation
	$$\Omega={\bar r}-{1\over 6}B{\bar r}^3-{1\over 8}K{\bar r}^4
	\quad .\eqno{(13)}$$
	Hence
	$$d\Omega=\big ( 1-{B\over 2}{\bar r}^2-{1\over 2}K{\bar r}^3 \big )
	d{\bar r}-{1\over 6}{\bar r}^3\sum_{i\neq 1}{{\partial B}\over
	{\partial x^i}}dx^i+{\cal O}({\bar r}^4)\quad .\eqno{(14)}$$
	In addition to eliminating $B$ and $K$, this changes three other metric
	coefficients, the changes being effected by
	$$\eqalign{&U\longrightarrow {\bar U}=U -{1\over 6} {{\partial B}\over
	{\partial t}}\quad ,\cr
	&V\longrightarrow{\bar V}=V-{1\over 6}{{\partial B}\over {\partial \theta}}
	\quad ,\cr
	&W\longrightarrow{\bar W}=W-{1\over 6}{{\partial B}\over {\partial \phi}}
	\quad .\cr}\eqno {(15)}$$

	Furthermore, we may make the coordinate transformation$^{(15)}$
	\def\baro{\bar r}
	\def\barx{\bar x}
	\def\bart{\bar t}
	\def\barh{\bar \theta}
	\def\barp{\bar \phi}
	$$x^i=\barx^i-{1\over 3}\baro^3F^i(\barx^j)-{1\over 4}\baro^4G^i(\barx^j)
	\quad ,\eqno{(16)}$$
	where $x^i\in \{ t,\theta,\phi\}$ and ${\bar x}^i=\{ \bart,\barh,\barp\}$.
	%Then we have
	%$$\eqalignno{dx^idx^j&=d\barx^id\barx^j+{\cal O}(\baro^2)\quad{\rm for}
	%\quad i\neq j,&{\rm (17a)}\cr
	%dx^id\baro&=d\barx^id\baro+{\cal O}(\baro^2)\quad ,&{\rm (17b)}\cr
	%dt^2=&d\bart^2-2\baro^2F^0(\barx^i)d\bart d\baro
	%-2\baro^3G^0(\barx^i)d\bart d\baro\cr
	%&\quad-{2\over 3}\baro^3\bigg ( {{\partial F^0}\over {\partial \bart}}
	%d\bart^2 + {{\partial F^0}\over {\partial \barh}}d\bart d\barh
	%+{{\partial F^0}\over {\partial \barp}} d\bart d\barp \bigg ) +{\cal O}
	%(\baro^4)\quad, &{\rm (17c)}\cr
	%d\theta^2=&d\barh^2-2\baro^2F^2(\barx^i)d\baro d\barh -2\baro^3G^2(\barx^i)
	%d\baro d\barh\cr
	%&\quad -{2\over 3}\baro^3\bigg ( {{\partial F^2}\over {\partial\bart}}
	%d\bart d\barh +{{\partial F^2}\over {\partial \barh}}d\barh^2
	%+{{\partial F^2}\over {\partial \barp}}d\barh d\barp \bigg )
	%+{\cal O}(\baro^4)
	%\quad ,&{\rm (17d)}\cr
	%d\phi^2=&d\barp^2-2\baro^2F^3(\barx^i)d\baro d\barp-2\baro^3G^3(\barx^i)
	%d\baro d\barp\cr
	%&\quad -{2\over 3}\baro^3\bigg ( {{\partial F^3}\over {\partial\bart}}
	%d\bart d\barp + {{\partial F^3}\over {\partial\barh}}d\barh d\barp
	%+{{\partial F^3}\over {\partial \barp}} d\barp^2\bigg )+{\cal O}(\baro^4)
	%\quad .&{\rm (17e)}\cr}$$
	This produces the conformal metric
	$$\eqalign{d{\tilde s}^2=&-d\bart^2+{{d\baro^2}\over {(1-\baro^2)}} +(1-
	\baro^2)(d\barh^2+\sin^2\barh d\barp^2)\cr
	&+\baro^2 \big ( Ad\bart^2+Cd\barh^2+D\sin^2\barh d\barp^2\big ) \cr
	&+2\baro^2 \big ( Ld\bart d\barh + Od\bart d\barp +Gd\barh d\barp\big ) \cr
	&+2\baro^2 \big [ (X+F^0) d\bart+(Y-F^2)d\barh+(Z-F^3\sin^2\barh)d\barp\big ]
	d\baro\cr
	&+\baro^3\bigg [ \bigg ( J+{2\over 3}{{\partial F^0}\over {\partial\bart}}
	\bigg )
	d\bart^2 +\bigg ( N-{2\over 3}{{\partial F^2}\over {\partial\barh}}\bigg )
	d\barh^2 \cr
	&\qquad\qquad+\bigg (F-{2\over 3}
	{{\partial F^3}\over {\partial\barp}}-{2\over 3}
	F^2\cot\barh\bigg )\sin^2\barh d\barp^2\bigg ]\cr
	&+2\baro^3\bigg [ \bigg ( M+{1\over 3}{{\partial F^0}\over {\partial\barh}}
	-{1\over 3}{{\partial F^2}\over {\partial\bart}}\bigg ) d\bart d\barh
	+\bigg ( S+{1\over 3}{{\partial F^0}\over {\partial\barp}}-{1\over
3}{{\partial
	F^3}\over {\partial \bart}}\sin^2\barh\bigg ) d\bart d\barp\cr
	&\qquad\qquad +\bigg ( Q-{1\over 3}
	{{\partial F^2}\over {\partial\barp}}-{1\over 3}{{\partial F^3}\over {\partial
	\barh}}\sin^2\barh\bigg ) d\barh d\barp\bigg ]\cr
	&+2\baro^3\big [ ({\bar U}+G^0)d\bart+({\bar V}-G^2)d\barh+({\bar W}-G^3\sin^2
	\barh)d\barp\big ]d\baro +{\cal O}(\baro^4)\quad ,\cr}\eqno{(17)}$$
	where the arguments of all functions appearing in this metric are the
	barred coordinates $\barx^i\in\{ \bart,\barh,\barp\} $. We choose the
	functions $F^i$ and $G^i$ to satisfy
	$$\eqalign{0=&X+F^0=Y-F^2=Z-F^3\sin^2\barh\cr
	=&{\bar U}+G^0={\bar V}-G^2={\bar W}-G^3\sin^2\barh\quad ,\cr}\eqno{(18)}$$
	and define
	$$\eqalign{{\bar M}=&M+{1\over 3}{{\partial F^0}\over {\partial\barh}}
	-{1\over 3}{{\partial F^2}\over {\partial\bart}}\cr
	=&M-{1\over 3}{{\partial X}\over {\partial\barh}}-{1\over 3}{{\partial Y}\over
	{\partial\bart}}\quad ,\cr}\eqno{(19)}$$
	and likewise make the obvious definitions for ${\bar S}$, ${\bar Q}$,
	${\bar J}$, ${\bar N}$, and ${\bar F}$.
	After this, the conformal metric simplifies to
	$$\eqalign{d{\tilde s}^2=&-d\bart^2+{{d\baro^2}\over
{(1-\baro^2)}}+(1-\baro^2)
	(d\barh^2+\sin^2\barh d\barp^2)\cr
	&+\baro^2(Ad\bart^2+Cd\barh^2+D\sin^2\barh d\barp^2)\cr
	&+2\baro^2(Ld\bart d\barh+Od\bart d\barp+Gd\barh d\barp)\cr
	&+\baro^3({\bar J}d\bart^2+{\bar N}d\barh^2+{\bar F}\sin^2\barh d\barp^2)\cr
	&+2\baro^3({\bar M}d\bart d\barh+{\bar S}d\bart d\barp+{\bar Q}d\barh d\barp)
	+{\cal O}(\baro^4)\quad ,\cr}\eqno{(20)}$$
	while the physical metric is
	$$ds^2=\Omega^{-2}d{\tilde s}^2\quad .\eqno{(21)}$$

	The next step is to compute the left-hand-side of the field equation (8)
	using the conformal metric (20). This task was performed both by hand and by
	the RCLASSI computer program for relativity calculations.
	The resulting field equations are
	$$\eqalignno{{4\over 3}A-B-{1\over 3}C-{1\over 3}D
	+{\cal O}(\baro)&=8\pi{\cal T}_{00}\quad ,&{\rm (22a)}\cr
	L+{\cal O}(\baro)&=8\pi{\cal T}_{02}\quad ,&{\rm (22b)}\cr
	O+{\cal O}(\baro)&=8\pi{\cal T}_{03}\quad ,&{\rm (22c)}\cr
	{2\over 3}(A-C-D)-B +{\cal O}(\baro)
	&=8\pi{\cal T}_{11}\quad ,&{\rm (22d)}\cr
	-{1\over 3}A+B+{4\over 3}C+{1\over 3}D+{\cal O}(\baro)
	&=8\pi{\cal T}_{22}\quad ,&{\rm (22e)}\cr
	G+{\cal O}(\baro)&=8\pi{\cal T}_{23}\quad ,&{\rm (22f)}\cr
	\big ( -{1\over 3}A+B+{1\over 3}C+{4\over 3}D\big )\sin^2\barh
	+{\cal O}(\baro)&=8\pi{\cal T}_{33}
	\quad ,&{\rm (22g)}\cr}$$
	where we define
	$${\cal T}_{ab}=g_{ac}{\cal T}^c_b
	=\Omega^{-2}{\tilde g}_{ac}{\cal T}^c_b
	={\cal O}(\Omega)={\cal O}(\baro)\quad ,\eqno{(23)}$$
	according to the specified fall-off condition (iii). Three components of the
	field equations are not written above because their left-hand-sides
	had no ${\cal O}(1)$ contributions.
	We see that we must have
	$$L=O=G=0\quad ,\quad A={1\over 2}B\quad ,\quad C=D=-{1\over 2}B\quad ,
	\eqno{(24)}$$
	and hence the conformal metric may be written as
	$$\eqalign{d{\tilde s}^2=&-d\bart^2+{{d\baro^2}\over
{(1-\baro^2)}}+(1-\baro^2)
	(d\barh^2+\sin^2\barh d\barp^2)\cr
	&+{B\over 2}\baro^2(d\bart^2-d\barh^2-\sin^2\barh d\barp^2)
	+\big ( Jd\bart^2+Nd\barh^2+F\sin^2\barh d\barp^2\cr
	&\qquad+2Md\bart d\barh +2Sd\bart d\barp +2Qd\barh
	d\barp \big )\baro^3 +{\cal O}(\baro^4)\quad .\cr}\eqno{(25)}$$
	If we let
	\def\barx{\bar \xi}
	$$d{\tilde s}_{\rm AdeS}^2=-d\bart^2+d\barx^2+\sin^2\barx
	(d\barh^2+\sin^2\barh
	d\barp^2)\quad ,\eqno{(26)}$$
	then the conformal metric can be written as
	$$\eqalign{d{\tilde s}^2=&\big ( 1-{1\over 2}B\cos^2\barx\big )
	d{\tilde s}^2_{\rm AdeS}
	+{1\over 2}B\cos^2\barx d\barx^2\cr
	&+\cos^3\barx\big ( {\bar J}d\bart^2+{\bar N}d\barh^2
	+{\bar F}\sin^2\barh d\barp^2\big )\cr
	&+2\cos^3\barx \big ( {\bar M}d\bart d\barh+{\bar S}d\bart d\barp
	+{\bar Q}d\barh d\barp\big )\cr
	&+{\cal O}(\cos^4\barx)\quad .\cr}\eqno{(27)}$$

	Purely as an aside, we take the opportunity to recalculate the field equations
	using the metric (27) in order to examine the ${\cal O}(\baro)$ terms.
	They reduce to the statements that
	${\cal T}_{\mu\nu}={\cal O}(\baro^2)$ for $\mu\neq\nu$,
	and that each of the diagonal
	components of ${\cal T}_{ab}$ are proportional to the combination
	$({\bar J}-{\bar N}-{\bar F}-{\bar K})\baro$,
	up to correction terms which are ${\cal O}(\baro^2)$.
	Thus, all the off-diagonal components of ${\cal T}_{ab}$ vanish at this
	order, and all the diagonal components will as well provided the trace
	vanishes, no doubt a consequence of the asymptotic symmetry.
	Ashtekar and Magnon alluded to the fact that the Bianchi identities
	and the ${\cal O}(\baro)$ fall-off condition on ${\cal T}_{ab}$ is strong
	enough to imply that many of the components of ${\cal T}_{ab}$ actually
	fall off as ${\cal O}(\baro^2)$. Note, however, that we do not
	require any condition on ${\bar J}-{\bar N}-{\bar F}-{\bar K}$ in what
follows,
	and so the diagonal components of ${\cal T}_{ab}$ need not vanish
	at ${\cal O}(\baro)$ here; we do not make any use of the ${\cal O}(\baro)$
	field equations.

	The Weyl tensor of this metric, to ${\cal O}(\baro)$ inclusive, is
	$$\matrix{&C_{0101}=-({\bar J}+{1\over 2}{\bar N}+{1\over 2}{\bar F})
	\baro\quad ,&C_{0112}={3\over 2}{\bar M}\baro\quad ,\cr & & \cr
	&C_{0113}={3\over 2}{\bar S}\baro\quad ,
	&C_{0202}=({\bar F}+{1\over 2}{\bar J}-{1\over 2}{\bar N})\baro\quad ,\cr
	& & \cr
	&C_{0203}=-{3\over 2}{\bar Q}\baro\quad ,
	&C_{0223}=-{3\over 2}{\bar S}\baro\quad ,\cr & & \cr
	&C_{0303}=({\bar N}+{1\over 2}{\bar J}-{1\over 2}{\bar F})
	\baro\sin^2\barh\quad ,&C_{0323}={3\over 2}{\bar M}\baro
	\sin^2\barh\quad ,\cr & & \cr
	&C_{1212}=({1\over 2}{\bar F}-{1\over 2}{\bar J}-{\bar N})\baro\quad ,
	&C_{1213}=-{3\over 2}{\bar Q}\baro\quad ,\cr & & \cr
	&C_{1313}=({1\over 2}{\bar N}-{1\over 2}{\bar J}-{\bar F})\baro
	\sin^2\barh\quad ,&C_{2323}=({1\over 2}{\bar F}+{1\over 2}{\bar N}+{\bar J})
	\baro\sin^2\barh
	\quad ,\cr}\eqno{(28)}$$
	with the other independent components being zero at this order.
	Since there is no component $C_{1ijk}$ at this order (where $i,j,k\in\{ 0,2,3
	\}$), the (rescaled) magnetic part of the Weyl tensor vanishes on ${\cal I}$.
	According to Ashtekar and Magnon, this is precisely the statement that the
	metric satisfies Hawking's ``reflexive'' boundary condition. Furthermore,
	note that the (rescaled) electric part of the Weyl tensor, defined by
	$$E_{cd}=\Omega^{-1}C_{acbd}n^a n^b\quad ,\quad n^a={\tilde g}^{ab}{\tilde
	\nabla}_b\Omega\quad ,\eqno{(29)}$$
	has its non-vanishing components on ${\cal I}$ given as
	$$\matrix{
	&E_{00}{\hat =}-({\bar J}+{1\over 2}{\bar N}+{1\over 2}{\bar F})\quad ,
	&E_{02}{\hat =}-{3\over 2}{\bar M}\quad ,\cr & & \cr
	&E_{22}{\hat =}{1\over 2}{\bar F}-{1\over 2}{\bar J}-{\bar N}\quad ,
	&E_{03}{\hat =}-{3\over 2}{\bar S}\quad ,\cr & & \cr
	&E_{33}{\hat =}({1\over 2}{\bar N}-{1\over 2}{\bar J}-{\bar F})
	\sin^2\barh\quad ,
	&E_{23}{\hat =}-{3\over 2}{\bar Q}\quad .\cr}\eqno{(30)}$$
	There cannot, of course, be any
	non-zero $E_{1\mu}$ component.

	Given the above construction, we may state the following result.

	{\narrower\smallskip\noindent
	{\underbar{Proposition 1:}}
	If $({\cal M},g_{ab})$ is asymptotically anti-de Sitter, there exist
	coordinates $(t,\xi,\theta,\phi)$ on ${\cal U}\cup{\cal I}$, where
	${\cal I}\subseteq{\bar {\cal U}}$ (the closure of ${\cal U}$)
	and ${\cal I}$ is the surface $\xi={{\pi}
	\over 2}$, in which the metric is conformal to
	$$\eqalign{
	d{\tilde s}^2=&\bigg ( 1-{1\over 2}B\cos^2\barx-{1\over 3}
	({\bar J}-{\bar N}-{\bar F})\cos^3\barx
	\bigg ) d{\tilde s}^2_{\rm AdeS}\cr
	&\qquad+\bigg ( {1\over 2}B
	+{1\over 3}({\bar J}-{\bar N}-{\bar F})\cos\barx\bigg )
	\cos^2\barx\ d\barx^2\cr
	&\qquad-{2\over 3}(\cos^3\barx) E_{ij} d{\bar x}^id{\bar x}^j
	+{\cal O}(\cos^4\barx)
	\quad ,\cr}\eqno{(31)}$$
	with $d{\tilde s}^2_{\rm AdeS}$ given by (26),
	and where the $E_{ij}$ are the components
	in this coordinate system
	of the electric part of the Weyl tensor (rescaled by $\Omega^{-1}$) on
	${\cal I}$, extended into spacetime by Lie dragging along lines of constant
	$\bart$, $\barh$, and $\barp$.\smallskip}

	\noindent
	The deviation of the metric from the anti-de Sitter metric is described at
	leading order essentially by the electric components of the Weyl tensor,
	which is natural because of the absence of gravitational radiation at
infinity.
	The above metric suffers from the usual coordinate singularities of the polar
	coordinates $(\barh,\barp)$ of course, so at least two charts are needed to
	cover ${\cal U}$ with {\it good} coordinates --- this is an artifact of the
	$S^2\times R$ topology of ${\cal I}$. However, this sort of difficulty is so
	mild that it is convenient simply to ignore it in what follows.

	\vskip 0.5 true cm
	\noindent
	{\centreline{\bf The Time Delay Formula}}
	\vskip 0.3 true cm
	\noindent
	Consider now the conformal metric. Along any null curve, not necessarily
	geodesic, we have
	$$\eqalign{0=&{{d{\tilde s}}\over {1-H\cos^2\xi}}\cr
	=&d{\tilde s}_{\rm AdeS}^2+H\cos^2\xi d\xi^2-{2\over 3}(\cos^3\xi)E_{ij}
	dx^idx^j+{\cal O}(\cos^4\xi)\quad ,\cr}\eqno{(32)}$$
	where
	$$H={1\over 2}B+{1\over 3}\big ( J-N-F\big ) \cos\xi\quad .\eqno{(33)}$$
	Throughout this section and the next,
	we drop the bars over the coordinates and over the functions
	$J$, $N$, {\it etc.}, and so (32) implies that
	$$\eqalign{0=&-dt^2+d\xi^2+\sin^2\xi (d\theta^2+\sin^2 d\phi^2)
	+H\cos^2\xi d\xi^2\cr
	&\qquad-{2\over 3}
	(\cos^3\xi) E_{\mu\nu}dx^{\mu}dx^{\nu}+{\cal O}(\cos^4\xi)\quad ,\cr}
	\eqno{(34)}$$
	whence of course
	$$dt^2=d\sigma^2+H\cos^2\xi d\xi^2
	-{2\over 3}(\cos^3\xi) E_{\mu\nu}dx^{\mu}dx^{\nu}+{\cal O}
	(\cos^4\xi)\quad ,\eqno{(35)}$$
	where
	$$d\sigma^2=d\xi^2+\sin^2\xi (d\theta^2+\sin^2\theta d\phi^2)\quad ,
	\eqno{(36)}$$
	is the ``round metric'' on $S^3$. We are not interested in the whole 3-sphere,
	but rather a region
	within the hemisphere $\xi\in [0,{{\pi}\over 2}]$, which represents a
	spatial slice of anti-de Sitter space, with $\xi={{\pi}\over 2}$ being
	infinity.

	Now consider the ``neighbourhood of infinity'' ${\cal U}$
	on which we are working. It
	is not $S^3\times R$, because it contains neither points $\xi\ge {{\pi}\over
	2}$ nor points $\xi\le {{\pi}\over 2}-\epsilon$, for some $\epsilon>0$.
	Rather, the neighbourhood can be described as
	${\cal U}={\cal T}\times R$, where
	${\cal T}$ is the 3-dimensional manifold $\{ x | \xi\in ({{\pi}\over 2}
	-\epsilon,{{\pi}\over 2}), \forall\theta,\forall\phi\}$; essentially
	${\cal T}$ is a 3-dimensional version of the Tropic of Capricorn (taking
	$\xi$ to be co-latitutude defined using the South Pole --- see figs. (2,3)).
	Then $\sigma$ defines an arc length on ${\cal T}$ with respect to the
	metric (36). The idea is to use this arc length
	to parametrise null curves in ${\cal Q}={\cal U}\cup{\cal I}$. Along such
	curves, we have
	$$\eqalign{dt=&\pm\bigg [ 1+H \big ( {{dr}\over {d\sigma}}\big )^2\cos^2\xi
	-{2\over 3}(\cos^3\xi) E_{\mu\nu}{{dx^{\mu}}\over
	{d\sigma}}{{dx^{\nu}}\over {d\sigma}}+{\cal O}(\cos^4\xi)\bigg ]^{1/2}
	d\sigma\cr
	=&\pm\bigg [ 1+{1\over 2}H \big ( {{dr}\over {d\sigma}}\big )^2\cos^2\xi
	-{1\over 3}(\cos^3\xi) E_{\mu\nu}{{dx^{\mu}}\over {d\sigma}}
	{{dx^{\nu}}\over {d\sigma}}+{\cal O}(\cos^4\xi)\bigg ] d\sigma\quad .\cr}
	\eqno{(37)}$$
	In particular, an arbitrary $C^0$ and piecewise $C^1$
	future-null curve that leaves $p\in{\cal I}$ and
	arrives at $q\in{\cal I}$, having traversed arc length $\sigma_{pq}$, will
	reach its destination at coordinate time
	$$t_q=t_p+\sum_{i=0}^n
	\int\limits^{\sigma_{i+1}}_{\sigma_i} \bigg ( 1+ {1\over 2}(\cos^2\xi)H
	\bigg ( {{dr}\over {d\sigma}}
	\bigg )^2-{1\over 3}(\cos^3\xi)
	E_{\mu\nu} {{dx^{\mu}}\over {d\sigma}} {{dx^{\nu}}\over {d\sigma}}
	+{\cal O}(\cos^4\xi)\bigg )d\sigma\quad ,\eqno{(38)}$$
	assuming $n$ points at whicch the curve fails to be differentiable.

	Let $\gamma$ be a null geodesic ruling ${\cal I}$ from $P$ to
	$Q$.\footnote{*}{It may seem surprising at first, but indeed this timelike
	${\cal I}$ admits null generators which are geodesics of $d{\tilde s}^2$--- in
	fact, it admits many
	families of them, about which more will be said shortly. Indeed, since
	equation (40) will show the {\it first-order} (but clearly not third-order)
	stability of the arrival time
	of $\gamma$ under the variation discussed here, the methods of ref. (16)
	verify that $\gamma$ is geodesic, given that it is null and lies
	on ${\cal I}$.}
	We vary this curve according to a very particular set of
	conditions, similar to those of ref. (16):
	\item{(i)}{The varied curves $\gamma'$ start at $P$ and end on the same
	generator of ${\cal I}$ ({\it i.e.} the same
	integral curve of ${{\partial}\over {\partial t}}$
	on ${\cal I}$) as contains $Q$, but they reach this generator at points
	$q$, where $t_q$ need not equal $t_Q$.}
	\item{(ii)}{The varied curves remain null in $d{\tilde s}^2$ and are $C^0$,
	piecewise $C^1$, so the
	{\it time-of-flight} formula (38) applies to all these curves.}
	\item{(iii)}{Each varied curve projects to a geodesic on $S^3$ with the metric
	$d\sigma^2$ --- hence they
	each traverse arc length $\pi$ as measured by $\sigma$.}
	\item{(iv)}{Along any varied curve $\gamma'$,
	if we denote ${\gamma'}^{\mu}(\sigma)=
	{{dx^{\mu}(\sigma)}\over {d\sigma}}$ as giving the tangent components and
	if we denote $\gamma^{\mu}(\sigma)$ as the tangent components for $\gamma$ at
	the same parameter value $\sigma$, then}
	$${\gamma'}^{\mu}(\sigma)=\gamma^{\mu}(\sigma)+{\cal O}(\cos\xi)\quad .
	\eqno{(39)}$$
	\item{(v)}{All varied curves remain within the region ${\cal Q}$, whence
	along each curve we have
	$\xi(\sigma)\in[{{\pi}\over 2}-\epsilon,{{\pi}\over 2}]$, $\forall\sigma$.}

	\noindent
	In particular, condition (iv) implies that $dr/d\sigma={\cal O}(\cos\xi)$,
	and so the {\it time-of-flight} along any curve obeying the above conditions
is
	$$t_q-t_P=\pi-{1\over 3}\cos^3\xi_0\int\limits^{\pi}_0
	\big ( E_{\mu\nu}\gamma^{\mu}\gamma^{\nu}+{\cal O}(\cos\xi)\big )
	d\sigma\quad ,\eqno{(40)}$$
	for some $\xi_0\in({{\pi}\over 2}-\epsilon,{{\pi}\over 2})$.
	The remaining
	integral is taken along the curve $\gamma\subseteq{\cal I}$, as this
	will induce only errors of ${\cal O}(\epsilon)$ in the integral, and
	need no longer be broken, as $\gamma$ is smooth. In other
	words, the integral in the following expression
	is to be taken along a curve lying on
	the metrical ({\it i.e.} round) 2-sphere $\xi={{\pi}\over 2}$.
	$$t_q-t_P=\pi-{1\over 3}\cos^3\xi_0\int\limits^{\pi}_0 \big (
	E_{\mu\nu}\gamma^{\mu}\gamma^{\nu}+{\cal O}(\epsilon)\big ) d\sigma\quad .
	\eqno{(41)}$$
	Finally, the {\it time-delay} (possibly negative) of the null curve $\gamma'$
	compared to that of $\gamma$ is
	$$\Delta t=(t_q-t_P)|_{\gamma'}-(t_Q-t_P)|_{\gamma}=-{1\over 3}\cos^3\xi_0
	\bigg ( \int\limits^{\pi}_0 E_{\mu\nu}\gamma^{\mu}\gamma^{\nu}d\sigma
	+{\cal O}(\epsilon)\bigg ) \quad .\eqno{(42)}$$
	For a pictorial description of the variation and the time delay effect,
	see figs. (2--4).

	Consider the case wherein the time-delay is
	negative. This will mean that the ``fastest curve'' from $P$ to a given
	timelike generator of ${\cal I}$ will not be one that remains always on
	${\cal I}$ --- it will pass through spacetime --- provided of course such
	a fastest curve exists. We will derive arguments why such a fastest curve
	cannot exist off ${\cal I}$, implying that the time delay is always
	non-negative. This implies a non-negativity condition on the integral in
	equation (42), which leads to a positive mass.

	\vskip 0.5 true cm
	\noindent
	{\centreline{\bf The Mass and Geodesic Focussing}}
	\vskip 0.5 true cm
	\noindent
	In this section, we will prove the following theorem:

	{\narrower\smallskip\noindent
	{\underbar {Theorem 2:}}
	Let $({\cal M}, g_{ab})$ be an asymptotically anti-de Sitter spacetime obeying
	the Borde energy condition and having a boundary-at-infinity ${\cal I}$
	such that, $\forall p\in{\cal I}$, there is a 2-sphere cross-section of
	${\cal I}$ that is not in the causal future of $p$.\footnote{**}
	{This condition prohibits serious causality violations which may be due
	entirely to curvature deep in ${\cal M}$ and therefore cannot be controlled
	by restrictions placed on the asymptotic properties of the spacetime.}
	Let $t$, parametrising the timelike Killing field
	${{\partial}\over {\partial t}}$ on ${\cal I}$, be normalised so that null
	geodesics on ${\cal I}$ with initial endpoint at $t=0$ reconverge at $t=\pi$,
	and let $\{ {\cal I}|t\in (T,T+\pi) \} = {\cal V}_T$.
	If $J^+({\cal V}_T)\cap J^-({\cal V}_T)$ is null geodesically complete, then
	the {\it average} mass on ${\cal V}_T$
	is non-negative. If the matter flux near ${\cal V}_T$
	vanishes as ${\cal O}(\Omega^5)$, the instantaneous mass is non-negative.
	\smallskip}

	First, we must say what we mean by the mass, the average
	mass, and the matter flux. Ashtekar and Magnon$^{(6)}$ define
	the following quantity associated with the timelike Killing field
	$\zeta^a={{\partial}\over {\partial t}}$ on ${\cal I}$.
	$$\mu=-{1\over {8\pi}}\int \Omega^{-1}C_{abcd}n^a n^c \zeta^b d^2S^d
	=-{1\over {8\pi}}\int E_{ab}\zeta^a d^2S^b
	\quad ,\eqno{(43)}$$
	where the integral is over a two-sphere cross-section of ${\cal I}$.
	Here $d^2S^a=t^adS^2$ where the measure $dS^2$
	is that of the induced metric on the 2-sphere (induced by the
	metric (1)), and $t^a$ lies in ${\cal I}$ and is a unit normal field for
	the 2-sphere (as computed using (1), of course).
	We may choose the cross-section orthogonal to ${{\partial}\over {\partial
t}}$,
	whence the integrand is $E_{00}$, the rescaled electric Weyl tensor.

	The conservation law follows from the Bianchi identities, which Ashtekar
	and Magnon showed to imply that
	$$D^aE_{ab}{\hat =}-{1\over 2} \lim_{\rightarrow{\cal I}}\Omega^{-4}
	{\cal T}_a^c q_{cb}n^a\quad ,\eqno{(44)}$$
	where $q_{ab}$ is the restriction of the conformal metric ${\tilde g}_{ab}$ to
	${\cal I}$ and $D_a$ is its compatible derivative. We may contract the free
	index with any conformal Killing field --- in this case we use
	$\zeta^a ={{\partial}\over {\partial t}}$ --- and integrate over a region
	$V\subseteq {\cal I}$ bounded by a pair of 2-sphere cross-sections ${\cal
	C}_1$ and ${\cal C}_2$ to obtain
	$$\bigg ( {{-1}\over {8\pi}}\bigg ) \int\limits_{{\cal C}_2} E_{ab}\zeta^a
	d^2S^b -\bigg ( {{-1}\over {8\pi}}\bigg ) \int\limits_{{\cal C}_1} E_{ab}
	\zeta^a d^2S^b = -{1\over 2}\int\limits_V \lim_{\rightarrow {\cal I}}
	\Omega^{-4} n^a{\cal T}^b_a q_{bc} \zeta^c = F_{\zeta}(V)\quad .\eqno{(45)}$$
	The integral $F_{\zeta}(V)$ describes the
	{\it matter flux} (or flux of material energy) at\footnote{$^{\dag}$}
	{The word ``at'' is incorrect, since the flux vanishes at ${\cal I}$,
	but it is perhaps descriptive. This integral actually describes flux through
	surfaces at arbitrarily large radii, and so perhaps ``near'' is a better
	word.}
	${\cal I}$ --- there is no gravitational wave (Bondi) flux.$^{(6)}$
	Let us choose
	the cross-sections ${\cal C}_{1,2}$ to be the surfaces $t_1=const.$ and
	$t_2=const.$ We define the ``mass at time $t$'' to be
	$$\mu(t)={{-1}\over {8\pi}}\int\limits_{{\cal C}(t)} E_{00} d^2S\quad .
	\eqno{(46)}$$
	The prefactor $-{1\over {8\pi}}$ normalises $\mu$ to equal the mass parameter
	$m$ for the AdeS-Schwarzschild metric.
	Then the conservation law becomes
	$$\mu(t_2)-\mu(t_1)=F_{\zeta}(V)\quad ,\eqno{(47)}$$
	where $V$ is the region of ${\cal I}$ wherein $t_1\le t\le t_2$.
	Lastly, we define
	$$\langle \mu \rangle = {1\over {\pi}}\int_T^{T+\pi}\mu(t) dt\eqno{(48)}$$
	to be the {\it average mass} on the interval $t\in [T,T+\pi]$.
	This period corresponds to that used to quotient Universal anti-de Sitter
	space to obtain anti-de Sitter space, with its closed timelike curves.

	We must also introduce the Borde energy condition,$^{(17,7)}$
	stated in Appendix 1. It is the
	weakest useful energy condition known to the author. Borde phrased his
	condition in terms of the Ricci tensor --- the field equations are used to
	obtain the formulation quoted
	in the appendix. As the theorem stated there refers
	only to null geodesics, the cosmological term does not appear --- the null
Borde
	condition is not sensitive to the trace of the stress-energy tensor --- which
	is a chief reason why much of this analysis passes over easily from ref. (8).
	Note the geodesic completeness condition is also weak; for example, it permits
	black holes and white holes.

	The plan of the proof is as follows. We will look for a ``fastest" causal
curve
	joining two antipodal
	timelike generating curves of ${\cal I}$ (antipodal generators are those that
	are separated by arc length $\pi$ in the parameter $\sigma$).
	A putative fastest curve
	is any null curve that lies completely
	on ${\cal I}$ and joins these two timelike
	generators. However, if the mass is negative,
	we will be able to construct a faster
	curve which joins these two timelike
	generators by leaving ${\cal I}$ and passing through spacetime, whence it
	follows that the fastest such curve passes through spacetime as well, the
	existence of such a fastest curve being guaranteed by geodesic completeness.
	Being fastest, it cannot focus ({\it i.e.} it cannot have a conjugate pair),
	in contradiction to the Borde theorem.\footnote{*}{For a discussion of
	conjugate pairs, geodesic completeness, achronal boundaries, and related
	issues necessary for what follows, see ref. (10). Also, see fig. (5), which
	shows null geodesics leaving the future boundary of some point to enter into
	the interior, where they ``focus'', developing locii of conjugate points
	called caustics.}
	The contradiction is
	avoided if the mass is non-negative on ${\cal I}$.

	The expression
	``$t$-generator'' will be used to refer to any one of the integral
	curves of ${{\partial}\over {\partial t}}$ which also happens to lie on
	${\cal I}$, since these curves generate ${\cal I}$. Now we are ready to prove
	the following proposition:

	{\narrower\smallskip\noindent
	{\underbar{Proposition 3:}}
	Given that the assumptions of Theorem 2 hold for ${\cal V}_T\subseteq {\cal
I}$,
	then
	$$\int\limits_0^{\pi}E_{ab}\gamma^a\gamma^bd\sigma\le 0$$
	along every null geodesic $\gamma$ ruling ${\cal V}_T$ and parametrised by
	$\sigma$ as defined above.\footnote{$^{\dag}$}{Note this quantity also plays
	a key role in the discussions of Ashtekar and Penrose$^{18}$ concerning
	ref. (7).}
	\smallskip}

	{\narrower\smallskip\noindent
	{\underbar{Proof:}}
	Let $P$ and $Q$ lie on antipodally opposed
	$t$-generators of ${\cal I}$,
	such that there
	is a null geodesic on ${\cal I}$ from $P$ to $Q$ (hence $P$ and $Q$ lie
	on the past and future boundaries, respectively, of ${\cal V}_T$).
	Because $P$ and $Q$ are on
	antipodal $t$-generators, the existence of one such geodesic implies the
	existence of a circle's worth (see
	fig.(4)), and there will be no faster curve on ${\cal I}$
	from $P$ to the $t$-generator containing $Q$, so we select any one of these
	null geodesics and call it $\gamma$. Now consider that $\int\limits_0^{\pi}
	E_{ab}\gamma^a\gamma^bd\sigma>0$ along $\gamma$.
	Then equation (42) for the time delay says that there is a null curve
$\gamma'$
	(one of the ones permitted by our variational procedure)
	that joins $P$ to the $t$-generator of $Q$ by traversing not ${\cal I}$
	but spacetime, which reaches that generator at
	some earlier point $Q'$, in virtue of having a negative time delay.

	In such a case,
	$Q$ will not lie on the {\it achronal boundary} set $\partial J^+(P)$, where
	$J^+(P)=J^+(P,{\tilde {\cal M}})$ denotes the set of points in $({\tilde
	{\cal M}},{\tilde g}_{ab})$ reached by future-causal curves from $P$.
	Say that this boundary meets $Q$'s $t$-generator at $Q''$ in the chronological
	past of $Q$. Such a $Q''$ must exist, by the given condition that ${\cal I}$
	contains a 2-sphere cross-section which lies outside the future of $P$.
	It is a standard theorem (see Section $6.3$ of ref. (10), for example)
	that one can trace back along the
	achronal boundary, obtaining a past-null geodesic, which has no endpoint
	except at $P$. This curve must enter the spacetime, for if it remained on
	${\cal I}$ it could not do better than $\gamma$, which joins $P$ to $Q$,
	not $Q''$. Furthermore, this curve must extend back to infinite negative
values
	of its affine parameter, for otherwise it would
	have segments that are
	past-inextendible (having no past endpoint) null geodesics
	in the spacetime metric $g_{ab}$ and yet are of finite affine length.
	This cannot happen because $J^+({\cal V}_T)\cap J^-({\cal V}_T)$ is null
	geodesically complete.\footnote{**}
	{The reader might object that $P$ isn't in ${\cal V}_T$, but is merely on
	$\partial {\cal V}_T$, and therefore we cannot assume geodesic completeness.
	Having established $P$ and $Q''$, we could, however, move the whole argument
	up a little, pushing $P$ and with it $Q''$ slightly into the future, but
	making sure they both remain to the past of $Q$. The objection is then
	overcome.}
	Because it reaches ${\cal I}$ in the future, and
	extends to negatively infinite affine parameter values in the past, and
	never leaves the achronal boundary $\partial J^+(P)$,
	this geodesic is an {\it infinite achronal geodesic} of the spacetime metric
	$g_{ab}$.

	However, the Borde theorem prohibits such a geodesic. Given the Borde energy
	condition, the Borde theorem guarantees that every infinite null geodesic has
a
	conjugate pair separated by a finite affine length; {\it i.e.} every
	null geodesic experiences the
	lensing effect of the geometry and is ``focussed''. Any null geodesic from $P$
	that has a conjugate
	pair cannot remain on the achronal boundary $\partial J^+(P)$ past the second
	of the two conjugate points of the pair, and therefore $\gamma''$ cannot have
a
	conjugate pair. We have established a contradiction.$\quad\dal$\smallskip}

	We note that the resolution of the contradiction, that $\int\limits^{\pi}_0
	E_{ab}\gamma^a\gamma^bd\sigma\le 0$ along each null geodesic $\gamma$ ruling
	${\cal I}$ from $P$ to $Q$ (and for each $P$ and $Q$ there are a
	circle's-worth of such $\gamma$), implies that the $\gamma$ themselves are
	``fastest'' curves (each being equally fast, of course). This is {\it not}
	in contradiction with the Borde theorem, of course, because these curves
	are not infinite achronal geodesics of the spacetime metric --- they are
	of only finite affine length in the conformal metric on ${\cal I}$ and,
	in any case, they encounter neither matter nor generic curvature which could
	cause them to focus.

	In order to prove Theorem 2, it will be useful to reiterate some facts
	concerning ${\cal I}$. While ${\cal I}$ can be generated by integral curves of
	${{\partial}\over {\partial t}}$,
	it can also be generated by any family
	of a one-parameter set of families of null geodesics tangent to ${\cal I}$
	(and indeed we have already made some use of these geodesics).
	Specifically, up to normalisation, a future-null
	vector field on ${\cal I}$ is given
	by specifying the parameter $\nu$ in
	$$\gamma^a={{\partial}\over {\partial t}}+\cos\nu {{\partial}\over {\partial
	\theta}}+{{\sin\nu}\over {\sin\theta}}{{\partial}\over {\partial\phi}}\quad .
	\eqno{(49)}$$
	Since ${\cal I}$ is a cylinder $R\times S^2$, we can fix a 2-sphere
	cross-section, say one of constant $t$, fix any $\nu$, and cover ${\cal I}$
	by dragging the 2-sphere along the resulting set of null integral
	curves.\footnote{$^{\dag}$}{In some formulations, ${\cal I}$ or, in actual
	fact, ${\cal V}_T$, is a hyperboloid.
	Then these null geodesic generators have an interesting historical
	connection. They form a circle's-worth of families of null geodesic
	generators of the 3-dimensional hyperboloid, and this is a direct
	generalisation of the observation, attributed$^{(19)}$ to the geometer and
	architect Sir Christopher Wren, that a 2-dimensional hyperboloid is
	generated by either of two families ({\it i.e.} an $S^0$'s-worth of
	families) of straight lines.}
	Note that of course each family $\nu=const.$ consists therefore of a
	2-sphere's-worth of null geodesics.

	It is useful for future purposes to compute
	the average over $\nu$ at a point of
	${\cal I}$ of $E_{ab}\gamma^a\gamma^b$.
	$$\eqalign{\int\limits_0^{2\pi}E_{ab}\gamma^a\gamma^bd\nu =
	&\int\limits_0^{2\pi}\big ( E_{00}+2E_{02}\cos\nu+2E_{03}{{\sin\nu}\over
	{\sin\theta}}+2E_{23}{{\sin\nu\cos\nu}\over {\sin\theta}}\cr
	&\qquad\qquad +E_{22}\cos^2\nu+E_{33}{{\sin^2\nu}\over {\sin^2\theta}}\big )
	d\nu\cr
	=&\int\limits^{2\pi}_0 \big (E_{00}+{1\over 2}E_{22}+{1\over {2\sin^2\theta}}
	E_{33}\big ) d\nu\cr
	=&\int\limits^{2\pi}_0 {3\over 2}E_{00} d\nu\cr
	=&3\pi E_{00}\quad ,\cr}\eqno{(50)}$$
	where we have used the tracelessness of $E_{ab}$ in an intermediate step.

	{\narrower\smallskip\noindent
	{\underbar {Proof of Theorem 2:}}
	Let ${\cal C}_0$ and ${\cal C}_1$ be two 2-sphere cross-sections of
	${\cal I}$, say they are the surfaces $t=t_0$ and $t=t_1>t_0$ respectively,
	and let them bound a region $V\subseteq{\cal I}$. Then
	$$\eqalign{\int\limits_V dV\int\limits_0^{2\pi} E_{ab}\gamma^a\gamma^b d\nu
	=&3\pi\int\limits_V E_{00} dV
	=3\pi\int\limits_{t_0}^{t_1} dt\int\limits_{{\cal C}(t)}
	E_{00}\sin\theta d\theta d\phi\cr
	=&-24\pi^2\int\limits_{t_0}^{t_1}\mu(t)dt\quad ,\cr}\eqno{(51)}$$
	where ${\cal C}(t)$ is the 2-sphere cross-section of ${\cal I}$ defined by
	$t_0\le t=const.\le t_1$
	and where $\mu(t)$ is the mass
	on that 2-sphere, as given by evaluating equation (43) or, equivalently, (46).

	Now rewrite the integral above making use of the null generators.
Specifically,
	let $t_0=T$ and choose $t_1$
	such that ${\cal C}_1$ can be reached from ${\cal C}_0$ by
	traversing a null generator from $\sigma=0$ to $\sigma=\pi$ (this distance
	is independent of the value of $\nu$ for the chosen null generator, since the
	surfaces ${\cal C}_0$ and ${\cal C}_1$ are surfaces of constant $t$). Then
	$V={\cal V}_T$ and
	$$\eqalign{\int\limits_{{\cal V}_T}
	dV \int\limits_0^{2\pi} E_{ab}\gamma^a\gamma^b d\nu
	=&\int\limits_0^{2\pi}d\nu \int\limits_{{\cal V}_T}
E_{ab}\gamma^a\gamma^bdV\cr
	=&\int\limits_0^{2\pi}d\nu\int\limits_{{\cal C}'} \sin\theta d\theta d\phi
	\int\limits_0^{\pi}E_{ab}\gamma^a\gamma^b d\sigma\quad ,\cr}\eqno{(52)}$$
	where ${\cal C}'$ is also a 2-sphere, but this time it is the 2-sphere of
	null generators for some fixed $\nu$.
	Now we have established in Proposition 3
	that the inner integral in equation (52)
	is bounded above by $0$ on the region of integration $\sigma\in [0,\pi]$,
	provided of course the stated assumptions hold on
	${\cal V}_T$. Writing
	this result as
	$$\int\limits_{0}^{\pi}E_{ab}\gamma^a\gamma^b d\sigma\le -b_{\gamma} \le 0
	\quad ,\eqno{(53)}$$
	where the bound $b_{\gamma}$ can depend on $\gamma$, then equation (52) gives
	$$\int\limits_{{\cal V}_T} dV \int\limits_0^{2\pi} E_{ab}\gamma^a\gamma^b d\nu
	\le\int\limits_0^{2\pi}d\nu\int\limits_{{\cal C}'}(-b_{\gamma})\sin\theta
	d\theta d\phi\le 0\quad .\eqno{(54)}$$
	By combining equations (46), (48), (50), and (54), we obtain
	$$\langle \mu \rangle=
	{1\over {\pi}}\int\limits_T^{T+\pi} \mu(t) dt
	\ge 0\quad .\eqno{(55)}$$
	This is the required result for the average mass.
	The result for the instantaneous mass derives from the conservation
	law (47) when the matter flux at ${\cal V}_T$ vanishes.
	$\quad\dal$\smallskip}

	\vskip 0.5 true cm
	\noindent
	{\centreline {\bf Conclusions}}
	\vskip 0.5 true cm
	\noindent
	The theorem is not a true positivity theorem in the usual sense.
	Setting aside for the moment the question of the time-averaged mass,
	we still have only a non-negativity
	theorem. In contrast to the hypersurface spinorial argument$^{(4)}$, it
	does not imply that exact AdeS spacetime is the {\it unique} ground
	state, leaving open the possibility of an instability of AdeS spacetime
	via quantum tunnelling between AdeS spacetime and
	another zero-energy state, perhaps one that obeys the Borde energy
	condition, but not the dominant energy condition, so as not to violate the
	theorem of ref. (4).

	It is  no surprise that the theorem refers to time-averaged mass when there
	is matter flux at ${\cal I}$. After all, the Borde energy condition also
	refers to time-averages of the matter tensor. It is quite reasonable that the
	consequence of having no null geodesic encounter net negative energy when
	averaged over its length (a weak paraphrasing of the Borde condition) is that
	the mass, averaged over time, should be positive. The time average is
	unnecessary in the case of asymptotically flat spacetimes because their null
	boundaries prevent what is entirely possible in the asymptotically AdeS case,
	that there can be inbound fluxes of matter energy near ${\cal I}$ which,
	by being present
	in sufficient quantity, change the sign of the total mass from what it was
	previously.

	Although the asymptotic structure used here was that of Universal AdeS
	spacetime (see part (ii) of Definition 1), the argument clearly goes
	through in any quotient spacetime whose universal cover is asymptotically
	AdeS in the sense of the Definition. If the covering spacetime admits an
	infinite achronal geodesic, as it will if the mass is negative,
	then this geodesic has no conjugate pair. It will project under the
	covering map to a null geodesic in the quotient spacetime, and this
	geodesic will also lack a conjugate pair (although it will no longer be
	achronal). It will therefore remain in contradiction with the Borde theorem.

	It is not possible to take the $\Lambda\rightarrow 0$ limit of the argument
	herein to obtain a positivity proof for the mass of an asymptotically flat
	spacetime. Such a limit may be described as singular, and several steps
	do not go through. The topology of ${\cal I}$ is changed significantly,
	due to the singularity at the point $i^0$ representing spatial infinity.
	As well, such a limit would fail to admit gravitational
	radiation at ${\cal I}^+$.

	However, it is very likely that methods quite similar to those discussed
	here will work in the other case, that of positive cosmological constant.
	This is often referred to as the asymptotically de Sitter case, but because
	the boundary-at-infinity for a de Sitter universe is comprised solely of
	two spacelike surfaces (called timelike infinity), the term usually
	refers instead to universes which asymptote to a
	positive spatial curvature Robertson-Walker metric that
	describes the region of de Sitter space inside of a cosmological
	horizon.$^{(9,10)}$ Ref. (9) discusses the McVittie metric, which describes
	an isolated mass embedded in a Robertson-Walker background spacetime. It
	is very straightforward to apply the method of ref. (11) to the McVittie
	case and relate positivity of the time delay to
	positivity of the McVittie mass parameter. I will
	return to this case in future work.

	Lastly, note the similarlity of the integral condition in Proposition 3
	governing the time delay (equivalently, the left-hand-side of equation (53))
	to the Averaged Null Energy Condition, {\it i.e.} to the Borde Condition
	${\rm (A.1.1)}$ in the case where the null geodesic path of integration
	is extended all the way to ${\cal I}$.\footnote{*}{I thank Gholamhossein
	Abolghasem for bringing this similarity to my attention.}
	It is conceivable that this may be suggestive of a quasi-local energy
	construction based on the electric components of Weyl defined with respect to
	and integrated over appropriate timelike 3-surfaces (which could be
	projected down to spacelike 2-surfaces when the flux across the surface
	vanished), and that such a construction might have certain desirable
	properties, such as positivity, at least when appropriate. We note that the
	idea that Weyl curvature describes in a quasi-local way the contribution of
	the gravitational field to the energy is a long-established one.$^{(20)}$

	\vskip 0.5 true cm
	\noindent
	{\centreline {\bf Acknowledgements}}
	\vskip 0,5 true cm
	\noindent
	I thank Malcolm MacCallum for communications concerning the
	use of the computer program RCLASSI, which was used to check equations
	(22) and to obtain equations (28).
	I thank the relativity and cosmology group of Dalhousie
	University for discussions and technical assistance with RCLASSI and
	with production of the figures, Abhay Ashtekar and Jorge Pullin for
	relevant correspondence, and my co-authors in ref. (7) for
	permission to reproduce one of the figures from it and for discussions on
	topics related to those appearing herein.
	\vskip 0.5 true cm
	\noindent
	{\centreline {\bf Appendix 1: Borde's Theorem}}
	\vskip 0.5 true cm
	\noindent
	Borde's paper$^{(17)}$ actually discusses focussing theorems for
	both timelike and null curves. Herein, we state a theorem for null curves
	alone, and we use the field equations to express the integral condition
	in terms of the matter tensor instead of the Ricci curvature.

	{\narrower\smallskip\noindent
	{\underbar{Borde's Focussing Theorem:}}
	Let $\gamma(t)$ be a complete affinely parametrised causal geodesic with
	tangent $\ell^a$ and let $\ell_{[a}R_{b|cd|e}\ell_{f]}\ell^c\ell^d\neq 0$
	somewhere on $\gamma$ (this is known as the null {\it generic condition}).
	Suppose that for any $\epsilon>0$ there is a $b>0$ such that for any
	$t_1<t_2$ there is a pair of intervals $I_1<t_1$ and $I_2>t_2$ of lengths
	$\ge b$ such that
	$$\int\limits_{t'}^{t''}T_{ab}\ell^a\ell^b dt \ge -\epsilon\quad \forall t'
	\in I_1,\forall t''\in I_2\quad .\eqno{\rm (A.1.1)}$$
	Then $\gamma$ contains a pair of conjugate points.\smallskip}

	\noindent
	In the text, the integral ${\rm (A.1.1)}$ is referred to as the Borde energy
	condition.
	\vskip 0.5 true cm
	\noindent
	{\centreline {\bf Appendix 2: The AdeS-Schwarzschild Solution}
	\vskip 0.5 true cm
	\noindent
	The AdeS-Schwarzschild solution is an exact solution whose metric can be
	written as
	$$\eqalign{ds^2=&-\big (
	1-{1\over 3}\Lambda r^2-{{2m}\over r}\big )dt^2 +{{dr^2}\over
	{\big (1-{1\over 3}\Lambda r^2 -{{2m}\over r}\big ) }}+r^2\big (d\theta^2+
	\sin^2\theta d\phi^2\big )\quad ,\cr
	=&-\big (1+r^2-{{2m}\over r}\big ) dt^2 +{{dr^2}\over {\big ( 1+r^2-{{2m}\over
	r}\big ) }}+r^2\big (d\theta^2+\sin^2\theta d\phi^2\big )\quad ,\cr}
	\eqno{\rm (A.2.1)}$$
	where we have rescaled $s$, $r$, $t$, and $m$ by $\sqrt{-3/\Lambda}$,
	recalling that of course $\Lambda<0$. We may now choose
	$$\Omega^2={1\over {1+r^2}}\quad .\eqno{\rm (A.2.2)}$$
	The metric becomes
	$$\eqalign{ds^2=
	&\Omega^{-2}\bigg \{ -dt^2+{{d\Omega^2}\over {1-\Omega^2}} +(1-\Omega^2)
	( d\theta^2+\sin^2\theta d\phi^2 ) \cr
	&\qquad\qquad+2m\Omega^3(dt^2+d\Omega^2)+{\cal O}(\Omega^4)\bigg \}\cr
	=&\Omega^{-2}d{\tilde s}^2\quad ,\cr}\eqno{\rm (A.2.3)}$$
	where $d{\tilde s}^2$ is the conformal metric. We read off $B=0$ and
	$K=2m$, so we perform the
	coordinate transformation
	$$\Omega=\baro-{1\over 4}m\baro^4\quad .\eqno{\rm (A.2.4)}$$
	Then the conformal metric becomes
	$$d{\tilde s}^2=-dt^2+{{d\baro^2}\over {1-\baro^2}}+(1-\baro^2)\big (
	d\theta^2+\sin^2\theta d\phi^2\big ) +2m\baro^3dt^2\quad .\eqno{\rm (A.2.5)}$$
	This can be written in the form
	$$\eqalign{d{\tilde s}^2=&\bigg ( 1-{2\over 3}m\baro^3\bigg ) \bigg [ -dt^2
	+{{d\baro^2}\over {1-\baro^2}}+(1-\baro^2)(d\theta^2+\sin^2\theta d\phi^2)
	\bigg ] \cr
	&\qquad +{2\over 3}m\baro^3 d\baro^2+{2\over 3}m\baro^3 \big ( 2dt^2
	+d\theta^2+\sin^2\theta d\phi^2\big ) +{\cal O}(\baro^4)\quad ,\cr}
	\eqno{\rm (A.2.6)}$$
	suggestive of equation (31).
	A simple computation (with RCLASSI) confirms that $E_{00}=-2m$, $E_{22}=-m$,
	and of course $E_{33}=-m\sin^2\theta$ for this metric. Thus, we recover
	the form (31). We may now transform to $\barx={\rm Arccos}\ \baro$
	and write
	$$d{\tilde s}^2=
	\bigg (1-{2m\over 3}\cos^3\barx \bigg ) d{\tilde s}_{\rm AdeS}^2
	-{2\over 3}(\cos^3\barx)E_{ij}dx^idx^j+{2m\over 3}\cos^3\barx d\barx^2
	+{\cal O}(\cos^4\barx)\quad .
	\eqno{\rm (A.2.7)}$$

	If $m>0$, then these
	metrics have event horizons which do not intersect ${\cal I}$.
	If $m<0$, no event horizons exist, and observers at
	${\cal I}$ can {\it always} see into the singularity, so a region ${\cal V}_T$
	obeying the geodesic completeness and other criteria
	required herein cannot be found.
	\vskip 0.5 true cm
	\noindent
	%\par\vfil\eject
	{\centreline{\bf References}}
	\item{(1)}{Breitenlohner, P., and Freedman, D.Z., {\it Ann. Phys.} {\underbar
	{144}}, 249 (1982).}
	\item{(2)}{Abbott, L.F., and Deser, S., {\it Nuc. Phys.} B{\underbar {195}},
	76 (1982).}
	\item{(3)}{Witten, E., {\it Commun. Math. Phys.} {\underbar {80}}, 381
(1981).}
	\item{(4)}{Gibbons, G.W., Hull, C.M., and Warner, N.P., {\it Nuc. Phys.}
	B{\underbar {218}}, 173 (1983).}
	\item{(5)}{Hawking, S.W., {\it Phys. Lett.} {\underbar {126}}B, 175
	(1983).}
	\item{(6)}{Ashtekar, A., and Magnon, A., {\it Class. Quantum Gravit.}
	{\underbar {1}}, L39 (1984).}
	\item{(7)}{Penrose, R., Sorkin, R.D., and Woolgar, E., preprint gr-qc/9301015
	(1993).}
	\item{(8)}{Gibbons, G.W., and Wells, C.G., DAMTP preprint R93/25 (1993),
	available as gr-qc/9310002.}
	\item{(9)}{Kastor, D., and Traschen, J., University of Massachusetts preprint
	UMHEP-399 (1993), available as gr-qc/9311025.}
	\item{(10)}{Hawking, S.W., and Ellis, G.F.R., {\it The Large Scale Structure
	of Spacetime} (Cambridge University Press, Cambridge, 1973).}
	\item{(11)}{Woolgar, E., preprint to appear in {\it Proc. Fifth Can. Conf.
	on Gen. Rel. and Rel. Astrophys.}, held at Waterloo, May 1993.}
	\item{(12)}{Chru\'sciel, P.T., MacCallum, M.A.H., and Singleton, D.B.,
	preprint gr-qc/9305021 (1993), and references therein.}
	\item{(13)}{Penrose, R., {\it Proc. Roy. Soc.} (London) A{\underbar {284}},
	159 (1965).}
	\item{(14)}{Wald, R.M., {\it General Relativity} (University of Chicago Press,
	Chicago, 1984).}
	\item{(15)}{Beig, R., and Schmidt, B.G., {\it Commun. Math. Phys.}
	{\underbar {87}}, 65 (1982).}
	\item{(16)}{See
	Nityananda, R., and Samuel, J., {\it Phys. Rev} D{\underbar {45}},
	3862 (1992) for a discussion of this variation and its application to Fermat's
	Principle in General Relativity.}
	\item{(17)}{Borde, A., {\it Class. Quantum Gravit.} {\underbar {4}}, 343
	(1987); see also Tipler, F.J., {\it J. Diff. Eq.} {\underbar {30}}, 165
(1978);
	{\it Phys. Rev.} D{\underbar {17}}, 2521 (1978).}
	\item{(18)}{Ashtekar, A., and Penrose, R., {\it Twistor Newsletter}
	{\underbar {31}}, 1 (1990).}
	\item{(19)}{Rouse Ball, W.W., {\it A Short Account of the History of
	Mathematics}, $4^{\rm th}$ ed. (1908) (reprinted by Dover Press, New York,
	1960).}
	\item{(20)}{Penrose, R., {\it Perspectives in Geometry and Relativity}
	(Hlavaty Festschrift), ed. B. Hoffmann, P. 259 (Indiana University Press,
	Bloomington, 1966).}
	\par\vfil\eject
	\noindent
	{\centreline{\bf Figure Captions}}
	\vskip 0.5 true cm
	\noindent
	\item{Fig. 1:}{Anti-de Sitter spacetime embedded in the Einstein cylinder.
	The ruled region is the image of the AdeS spacetime, while the vertical lines
	constitute the boundary ${\cal I}$. Universal AdeS spacetime is represented
	by an infinite vertical strip on this diagram.}
	\item{Fig. 2:}{$\gamma$ and $\gamma'$. Here $\gamma$ is ruling ${\cal I}$,
	while $\gamma'$ moves alittle into spacetime. The ``neighbourhood of
	infinity'' in which $\gamma'$ is constrained to move is represented by
	the region between the outer cylinder (${\cal I}$ itself) and the inner
	cylinder. In the case drawn here, the time delay of $\gamma'$ is
	positive.}
	\item{Fig. 3:}{This traces the progress of $\gamma$ and $\gamma'$ through
	space (actually, space plus boundary) at successive intervals of time. Space
	here is a hemisphere, which would be accurate for the exact AdeS case. The
	boundary (infinity) is the equator $\xi={{\pi}\over 2}$, and $\gamma$ moves
	along this equator while $\gamma'$ moves through spacetime near the equator.
	The coordinates $\xi$ and $\phi$ are shown; $\theta$ is suppressed.}
	\item{Fig. 4:}{This shows successive $t=const.$ slices of ${\cal I}$. Since
	${\cal I}=S^2\times R$, no coordinate is suppressed. From any point $p\in
	{\cal I}$, there originates a circle's-worth of null geodesics which rule
	${\cal I}$. Each of these geodesics traverses the $S^2$ and arrives at the
	antipodal point at the same time as all the others. This sequence of sketches
	tracks their progress.}
	\item{Fig. 5:}{A null cone of a point $p$
	in a generic curved spacetime on which the Borde
	energy condition holds. The null geodesics are deflected by the curvature
	and begin to converge upon conjugate points which lie along the caustics,
	which are the cuspoidal structures in the light cone's interior.
	They first pass through a ``crossing region'' and leave the boundary of the
	future of $p$.}
	\bye